\documentclass[twocolumn,aps,showpacs,floatfix,prc]{revtex4}
\usepackage[dvips]{epsfig}
\begin{document}
%\draft

\title{ Lambda hyperonic effect on the normal driplines }

\author{ C. Samanta$^{1,2,3}$, P. Roy Chowdhury$^1$ and D.N. Basu$^4$ }

\affiliation{ $^1$ Saha Institute of Nuclear Physics, 1/AF Bidhan Nagar, Kolkata 700 064, India }
\affiliation{ $^2$Physics Department, Virginia Commonwealth University, Richmond, VA 23284-2000, U.S.A. }
\affiliation{ $^3$Physics Department, University of Richmond, VA 23173, U.S.A. }
\affiliation{ $^4$ Variable  Energy  Cyclotron  Centre, 1/AF Bidhan Nagar, Kolkata 700 064, India }

\email[E-mail : ]{chhanda.samanta@saha.ac.in; csamanta@richmond.edu}
\email[E-mail : ]{partha.roychowdhury@saha.ac.in} 
\email[E-mail : ]{dnb@veccal.ernet.in}

\date{\today }

\begin{abstract}

    A generalized mass formula is used to calculate the neutron and proton drip lines of normal and lambda hypernuclei treating non-strange and strange nuclei on the same footing. Calculations suggest existence of several bound hypernuclei whose normal cores are unbound. Addition of $\Lambda$ or, $\Lambda\Lambda$ hyperon(s) to a normal nucleus is found to cause shifts of the neutron and proton driplines from their conventional limits.
\end{abstract}

\noindent

\maketitle

\noindent
\section{Introduction}

Knowledge of baryon-baryon interaction under flavored-SU(3) with up, down and strange quarks is important for the comprehensive understanding of the universe. 
For example, current understanding of the high density core of a neutron star is a substantial matter consisting of hyperons \cite{Mi03}. It has been predicted that in a certain range of parameters, formations of metastable hyperonic matter can lead to the twin star \cite{Bi02} and the collapse to a twin star might have similar properties of a hypernova collapse \cite{Ka83} leading to emission of gravitational waves. 

In a relativistic model, it was shown that a strong hyperon-hyperon interaction can lead to a vast array of stable objects composed of n, p, $\Lambda$, $\Xi^0$, $\Xi^-$ baryons of arbitrarily large baryon number A, very high strangeness content and small net charge~\cite{Sc93}. The maximal binding energy per baryon was predicted to be E$_B$/A $\sim$ -21 MeV with strangeness per baryon $f_s \sim$ 1-1.2, charge per baryon $f_q \sim$ -0.1 to 0 and baryon density 2.5-3 times that of ordinary nuclei. For A$\ge$6, stable combinations can be obtained involving only $\Lambda$, $\Xi^0$, $\Xi^-$ hyperons. These results can have important consequences in relativistic heavy ion collisions as well as in astrophysical sites.

While nucleon-nucleon interaction is well studied, very little is known about hyperon-hyperon interaction. Hyperons have very short life times. Hence the baryon-baryon interaction involving hyperon is usually studied through the production of hypernuclei. Hypernuclear physics projects with heavy ion, meson, electron and anti-proton beams have been undertaken in several facilities in Europe and Japan. Experiments have been planned to be carried out at HypHI at GSI, FINUDA at Frascati in Italy, PANDA at FAIR, KaoS at MAMI C in Mainz and Hyperball at KEK and J-PARC in Japan \cite{gsi}. First results on $^{12}C_{\Lambda}$ production at DA$\Phi$NE have been obtained with the FINUDA spectrometer \cite{Ag05}. 

Recently, Botvina and Pochodzalla \cite{Bo07} have suggested that multi-fragmentation of strange spectators produced in peripheral collisions of relativistic heavy ions offer a new possibility for investigating hypernuclei under conditions essentially different from those accessible in conventional nuclear structure studies. Production of hypernuclei and their properties were studied using two different mass formulas \cite{gr95,cs05,cs06}. It was shown that as the fragment yield is sensitive to binding energy of hypernuclei one can distinguish between different mass formulas of hypernuclei.  

The $\Lambda$ particle provides the nuclear core with additional binding and causes shrinkage of the hypernucleus. This shrinkage was predicted theoretically~\cite{hiyama} and confirmed experimentally to be of the order of $\sim$20$\%$~\cite{tanida}. Such confirmation was possible due to excellent energy resolution of the $gamma$-ray hypernuclear data. The relativistic Hartree-Bogoliubov model in co-ordinate space with finite range pairing interaction predicts that although the inclusion of the lambda hyperon to Ne-isotopes does not produce excessive changes in bulk properties, through a purely relativistic effect, it can increase the spin-orbit term which binds the outermost neutrons. This can shift the neutron drip line by stabilizing an otherwise unbound core nucleus \cite{Vr98}.

Some of the proposed experiments at GSI want to explore lambda hypernuclei with nucleonic cores far from the beta stability line as information on very neutron rich hypernuclei are essential to understand the nature of the neutron stars. In this context it is now pertinent to estimate the limits of binding of the neutron-rich and proton-rich lambda hypernuclei. In this work we investigate the above limits on the basis of the generalised mass formula (BWMH) developed by us earlier \cite{cs05,cs06}. BWMH mass formula (or, ``Samanta Formula" as named by Botvina and Pochodzalla~\cite{Bo07}) is applicable to both non-strange normal  nuclei and strange hypernuclei. It was formulated on the basis of the extended liquid drop model (BWM) of Samanta and Adhikari~\cite{Sa02} developed earlier for non-strange normal nuclei. A hyperon term (H) was added to BWM~\cite{cs05} to treat the normal nuclei and hypernuclei on the same footing while preserving the standard nuclear matter properties \cite{bc04}. BWMH was fitted to experimental data on non-strange normal nuclei as well as several strange hypernuclei~\cite{cs05,cs06} and a fixed set of parameters for all non-strange normal nuclei and strange hypernuclei was prescribed. Since this mass formula is based on a liquid drop model, any exotic feature due to sudden change of structure near the drip lines would not be reflected in the current investigation. Nevertheless, it is expected to provide a guideline for the future experiments with lambda and double lambda hypernuclei in the same spirit as the drip lines predicted for the normal non-strange nuclei~\cite{csb05}.

The generalised mass formula (BWMH)~\cite{cs05,cs06} is applicable to both the $\Lambda$ and $\Lambda\Lambda$ hypernuclei and it well reproduces the available experimental data \cite{Ba90,Ta01} on $\Lambda$ and $\Lambda\Lambda$ hyperon(s) separation energies. Here we use BWMH to compute the neutron and proton driplines of non-strange normal nuclei as well as strange $\Lambda$ and $\Lambda\Lambda$ hypernuclei. Calculations show that addition of a single Lambda shifts the normal neutron and proton driplines and addition of double Lambda makes the effect more prominent. A detailed study near the driplines reveals the possibility of several bound hypernuclei beyond the normal neutron- and proton- drip lines. 

\noindent
\section{Generalised mass formula (BWMH) and nuclear stability}

Stability of normal nuclei and lambda hypernuclei are studied using the mass formula BWMH \cite{cs06}. 
Lambda hypernuclei considered here have either one $\Lambda$ or two $\Lambda$ hyperon(s) added to a normal nucleus which is treated as a core. For each element of a fixed proton number (Z), the neutron number is varied until the last bound nuclei are found at the neutron-rich and neutron-deficient sides. The neutron and proton drip lines of Lambda hypernuclei found by this method are compared with the normal drip lines. The total binding energy of a hypernucleus of total mass number A and net charge Z containing charged or neutral hyperon(s) was given by the following equation \cite{cs06}:

\begin{eqnarray}
B(A,Z) = 15.777A-18.34A^{2/3}-0.71\frac{Z(Z-1)}{A^{1/3}} \nonumber\\
-\frac{23.21(N-Z_c)^2}{[(1+e^{-A/17})A]}+(1-e^{-A/30})\delta\ \nonumber\\
 + n_Y [0.0335(m_Y) - 26.7 - 48.7 \mid S \mid}{A^{-2/3}], 
\label{seqn1}
\end{eqnarray}
\noindent
where $\delta=12A^{-1/2}$ for $N,Z_c$ even, $=-12A^{-1/2}$ for $N,Z_c$ odd, = 0 otherwise, $n_Y$ = number of hyperons in a nucleus, $m_Y$ = mass of the hyperon in $MeV$, $S$ = strangeness of the hyperon and the mass number $A = N + Z_c + n_Y$ is equal to the total number of baryons. $N$ and $Z_c$ are the number of neutrons and protons respectively while the $Z$ in eqn.(1) is  given by $Z = Z_c + n_Y q$ where $q$ is the charge number (with proper sign) of hyperon(s) constituting the hypernucleus. For non-strange (S=0) normal nuclei, $Z_c = Z$ as $n_Y$ =0. The choice of $\delta$ value depends on the number of neutrons and protons being odd or even in both the cases of normal and hypernuclei. For example, in case of $^{13}_{\Lambda}C$, $\delta=+12A^{-1/2}$ as the (N, $Z_c$) combination is even-even, whereas, for non-strange normal $^{13}C$ nucleus $\delta=0$ as A=13(odd). 

The hyperon term (last term in equation 1) reflects SU(6) symmetry breaking through explicit consideration of the different masses of different hyperons. The three coefficients of the hyperon term were obtained by minimizing root mean square deviation of the theoretical hyperon separation energies from the experimental ones~\cite{Ba90,Maj95,Pile91,Lala88,Lala94,Hase96,Aoki91,Dov91}. The hyperon separation energy ($S_Y$) is defined as

\begin{equation}
S_Y = B(A,Z)_{hyper} - B(A-n_Y, Z_c)_{core},
\label{seqn2}
\end{equation}
\noindent
which is the difference between the binding energy of a hypernucleus and the binding energy of its non-strange core nucleus.\\

To study the hyperonic effect on the nuclear binding of the nuclei at the dripline it is necessary to study the single neutron $(S_n)$ and single proton $(S_p)$ separation energies for all isotopes of each element using BWMH. The comparison of the normal nuclear dripline with the hypernuclear dripline illustrates the effect of the modified nuclear dynamics at the dripline.
The neutron separation energies ($S_n$) and proton separation energies ($S_p$) from the hypernuclei containing a single $\Lambda$ or, $\Lambda\Lambda$ hyperon(s) are defined as   
\begin{equation}
S_n = B(A,Z)_{hyper} - B(A-1, Z)_{hyper},
\end{equation}

\begin{equation}
S_p = B(A,Z)_{hyper} - B(A-1, Z-1)_{hyper},
\label{seqn3 and 4}
\end{equation}
\noindent
where $B(A,Z)_{hyper}$ is the binding energy of a hypernucleus with a hyperon(s) inside with A and Z being the total mass number and net charge number respectively. Their respective binding energies provide necessary guidelines for studying the nuclear stability near the driplines against decay by emission of protons or neutrons. The neutron dripline is defined as the last bound neutron rich nuclei beyond which the neutron separation energy changes sign and becomes negative. Similarly the proton dripline nucleus is defined as the last bound neutron deficient nucleus beyond which the proton separation energy becomes negative.

\begin{table}[htbp]
\caption{One-nucleon separation energies (in MeV) on driplines for each element with the lowest and highest number of bound neutrons in normal and $\Lambda $-hypernuclei.}
\begin{ruledtabular}
\begin{tabular}{ccccc}
Symbol&Normal&Normal&Hyper&Hyper \\
&p-drip&n-drip&p-drip&n-drip \\ \hline 
$Z$&$^AZ$, $S_p$&$^AZ$, $S_n$&$^AZ$, $S_p$&$^AZ$, $S_n$ \\ \hline
\hline
${\bf Li}$&$^{  5} Li$,   3.36&$^{ 11} Li$,    .58&$^{  5}_{\Lambda}Li$,   1.22&$^{ 12}_{\Lambda}Li$,   1.87\\
${\bf Be}$&$^{  6} Be$,   1.11&$^{ 14} Be$,    .90&$^{  7}_{\Lambda} Be$,   3.79&$^{ 15}_{\Lambda} Be$,   1.84\\
${\bf B }$&$^{  8} B $,    .74&$^{ 17} B $,    .99&$^{  9}_{\Lambda} B $,   2.39&$^{ 18}_{\Lambda} B $,   1.73\\
${\bf C }$&$^{  9} C $,    .17&$^{ 20} C $,   1.01&$^{ 10}_{\Lambda} C $,   1.80&$^{ 21}_{\Lambda} C $,   1.63\\
${\bf N }$&$^{ 12} N $,   1.98&$^{ 23} N $,    .97&$^{ 12}_{\Lambda} N $,    .24&$^{ 24}_{\Lambda} N $,   1.50\\
${\bf O }$&$^{ 13} O $,   1.98&$^{ 26} O $,    .94&$^{ 13}_{\Lambda} O $,    .44&$^{ 27}_{\Lambda} O $,   1.40\\
${\bf F }$&$^{ 15} F $,    .09&$^{ 29} F $,    .89&$^{ 16}_{\Lambda} F $,    .81&$^{ 32}_{\Lambda} F $,    .01\\
${\bf Ne}$&$^{ 16} Ne$,    .64&$^{ 32} Ne$,    .87&$^{ 17}_{\Lambda} Ne$,   1.39&$^{ 35}_{\Lambda} Ne$,    .04\\
${\bf Na}$&$^{ 19} Na$,    .54&$^{ 35} Na$,    .84&$^{ 20}_{\Lambda} Na$,   1.05&$^{ 38}_{\Lambda} Na$,    .08\\
${\bf Mg}$&$^{ 20} Mg$,   1.37&$^{ 38} Mg$,    .84&$^{ 20}_{\Lambda} Mg$,    .05&$^{ 41}_{\Lambda} Mg$,    .13\\
${\bf Al}$&$^{ 23} Al$,    .71&$^{ 41} Al$,    .84&$^{ 24}_{\Lambda} Al$,   1.10&$^{ 44}_{\Lambda} Al$,    .18\\
${\bf Si}$&$^{ 23} Si$,    .07&$^{ 44} Si$,    .86&$^{ 24}_{\Lambda} Si$,    .56&$^{ 47}_{\Lambda} Si$,    .24\\
${\bf P }$&$^{ 27} P $,    .74&$^{ 49} P $,    .04&$^{ 28}_{\Lambda} P $,   1.06&$^{ 50}_{\Lambda} P $,    .31\\
${\bf S }$&$^{ 27} S $,    .46&$^{ 52} S $,    .12&$^{ 28}_{\Lambda} S $,    .86&$^{ 53}_{\Lambda} S $,    .37\\
${\bf Cl}$&$^{ 31} Cl$,    .68&$^{ 55} Cl$,    .21&$^{ 32}_{\Lambda} Cl$,    .95&$^{ 56}_{\Lambda} Cl$,    .44\\
${\bf Ar}$&$^{ 31} Ar$,    .69&$^{ 58} Ar$,    .29&$^{ 32}_{\Lambda} Ar$,   1.02&$^{ 59}_{\Lambda} Ar$,    .51\\
${\bf K }$&$^{ 35} K $,    .57&$^{ 61} K $,    .37&$^{ 36}_{\Lambda} K $,    .80&$^{ 62}_{\Lambda} K $,    .58\\
${\bf Ca}$&$^{ 35} Ca$,    .81&$^{ 64} Ca$,    .45&$^{ 36}_{\Lambda} Ca$,   1.08&$^{ 67}_{\Lambda} Ca$,    .07\\
${\bf Sc}$&$^{ 39} Sc$,    .42&$^{ 67} Sc$,    .53&$^{ 40}_{\Lambda} Sc$,    .62&$^{ 70}_{\Lambda} Sc$,    .17\\
${\bf Ti}$&$^{ 39} Ti$,    .84&$^{ 72} Ti$,    .07&$^{ 40}_{\Lambda} Ti$,   1.08&$^{ 73}_{\Lambda} Ti$,    .25\\
${\bf V }$&$^{ 43} V $,    .24&$^{ 75} V $,    .17&$^{ 44}_{\Lambda} V $,    .42&$^{ 76}_{\Lambda} V $,    .34\\
${\bf Cr}$&$^{ 43} Cr$,    .81&$^{ 78} Cr$,    .25&$^{ 44}_{\Lambda} Cr$,   1.02&$^{ 79}_{\Lambda} Cr$,    .41\\
${\bf Mn}$&$^{ 47} Mn$,    .04&$^{ 81} Mn$,    .34&$^{ 48}_{\Lambda} Mn$,    .20&$^{ 84}_{\Lambda} Mn$,    .05\\
${\bf Fe}$&$^{ 47} Fe$,    .73&$^{ 84} Fe$,    .41&$^{ 48}_{\Lambda} Fe$,    .92&$^{ 87}_{\Lambda} Fe$,    .13\\
${\bf Co}$&$^{ 52} Co$,    .76&$^{ 89} Co$,    .08&$^{ 53}_{\Lambda} Co$,    .89&$^{ 90}_{\Lambda} Co$,    .22\\
${\bf Ni}$&$^{ 51} Ni$,    .61&$^{ 92} Ni$,    .16&$^{ 52}_{\Lambda} Ni$,    .78&$^{ 93}_{\Lambda} Ni$,    .29\\
${\bf Cu}$&$^{ 56} Cu$,    .48&$^{ 95} Cu$,    .24&$^{ 57}_{\Lambda} Cu$,    .60&$^{ 98}_{\Lambda} Cu$,    .00\\
${\bf Zn}$&$^{ 55} Zn$,    .47&$^{ 98} Zn$,    .31&$^{ 56}_{\Lambda} Zn$,    .62&$^{101}_{\Lambda} Zn$,    .08\\
${\bf Ga}$&$^{ 60} Ga$,    .19&$^{103} Ga$,    .04&$^{ 61}_{\Lambda} Ga$,    .30&$^{104}_{\Lambda} Ga$,    .16\\
${\bf Ge}$&$^{ 59} Ge$,    .30&$^{106} Ge$,    .11&$^{ 60}_{\Lambda} Ge$,    .44&$^{107}_{\Lambda} Ge$,    .22\\
${\bf As}$&$^{ 65} As$,    .66&$^{109} As$,    .18&$^{ 65}_{\Lambda} As$,    .00&$^{110}_{\Lambda} As$,    .30\\
${\bf Se}$&$^{ 63} Se$,    .11&$^{112} Se$,    .25&$^{ 64}_{\Lambda} Se$,    .23&$^{115}_{\Lambda} Se$,    .05\\
${\bf Br}$&$^{ 69} Br$,    .33&$^{117} Br$,    .02&$^{ 70}_{\Lambda} Br$,    .42&$^{118}_{\Lambda} Br$,    .12\\
${\bf Kr}$&$^{ 68} Kr$,    .62&$^{120} Kr$,    .08&$^{ 68}_{\Lambda} Kr$,    .02&$^{121}_{\Lambda} Kr$,    .19\\
${\bf Rb}$&$^{ 73} Rb$,    .00&$^{123} Rb$,    .15&$^{ 74}_{\Lambda} Rb$,    .08&$^{124}_{\Lambda} Rb$,    .25\\
${\bf Sr}$&$^{ 72} Sr$,    .36&$^{126} Sr$,    .21&$^{ 73}_{\Lambda} Sr$,    .46&$^{129}_{\Lambda} Sr$,    .04\\
${\bf Y }$&$^{ 78} Y $,    .33&$^{131} Y $,    .01&$^{ 79}_{\Lambda} Y $,    .40&$^{132}_{\Lambda} Y $,    .10\\
${\bf Zr}$&$^{ 76} Zr$,    .10&$^{134} Zr$,    .07&$^{ 77}_{\Lambda} Zr$,    .19&$^{135}_{\Lambda} Zr$,    .16\\
${\bf Nb}$&$^{ 83} Nb$,    .58&$^{137} Nb$,    .13&$^{ 83}_{\Lambda} Nb$,    .04&$^{138}_{\Lambda} Nb$,    .22\\
${\bf Mo}$&$^{ 81} Mo$,    .45&$^{140} Mo$,    .19&$^{ 82}_{\Lambda} Mo$,    .53&$^{143}_{\Lambda} Mo$,    .03\\
${\bf Tc}$&$^{ 87} Tc$,    .20&$^{145} Tc$,   .01&$^{ 88}_{\Lambda} Tc$,  .26&$^{146}_{\Lambda} Tc$,  .09\\
%\noalign{\smallskip}
\end{tabular}
\end{ruledtabular}
\end{table}

\begin{table}[htbp]
\caption{One-nucleon separation energies (in MeV) on driplines for each element with the lowest and highest number of bound neutrons in normal and $\Lambda$-hypernuclei.}
\begin{ruledtabular}
\begin{tabular}{ccccc}
Symbol&Normal&Normal&Hyper&Hyper \\
&p-drip&n-drip&p-drip&n-drip \\ \hline 
$Z$&$^AZ$, $S_p$&$^AZ$, $S_n$&$^AZ$, $S_p$&$^AZ$, $S_n$ \\ \hline
\hline
${\bf Ru}$&$^{ 85} Ru$,    .14&$^{148} Ru$,    .07&$^{ 86}_{\Lambda} Ru$,    .22&$^{149}_{\Lambda} Ru$,    .15\\
${\bf Rh}$&$^{ 92} Rh$,    .40&$^{151} Rh$,    .13&$^{ 93}_{\Lambda} Rh$,    .45&$^{152}_{\Lambda} Rh$,    .20\\
${\bf Pd}$&$^{ 90} Pd$,    .39&$^{154} Pd$,    .18&$^{ 91}_{\Lambda} Pd$,    .46&$^{157}_{\Lambda} Pd$,    .03\\
${\bf Ag}$&$^{ 96} Ag$,    .01&$^{159} Ag$,    .02&$^{ 97}_{\Lambda} Ag$,    .06&$^{160}_{\Lambda} Ag$,    .09\\
${\bf Cd}$&$^{ 94} Cd$,    .06&$^{162} Cd$,    .07&$^{ 95}_{\Lambda} Cd$,    .12&$^{163}_{\Lambda} Cd$,    .14\\
${\bf In}$&$^{101} In$,    .14&$^{165} In$,    .12&$^{102}_{\Lambda} In$,    .18&$^{166}_{\Lambda} In$,    .19\\
${\bf Sn}$&$^{ 99} Sn$,    .25&$^{168} Sn$,    .17&$^{100}_{\Lambda} Sn$,    .30&$^{171}_{\Lambda} Sn$,    .04\\
${\bf Sb}$&$^{106} Sb$,    .24&$^{173} Sb$,    .03&$^{107}_{\Lambda} Sb$,    .28&$^{174}_{\Lambda} Sb$,    .09\\
${\bf Te}$&$^{104} Te$,    .38&$^{176} Te$,    .07&$^{105}_{\Lambda} Te$,    .43&$^{177}_{\Lambda} Te$,    .14\\
${\bf I }$&$^{111} I $,    .30&$^{179} I $,    .13&$^{112}_{\Lambda} I $,    .33&$^{182}_{\Lambda} I $,    .00\\
${\bf Xe}$&$^{108} Xe$,    .02&$^{182} Xe$,    .17&$^{109}_{\Lambda} Xe$,    .07&$^{185}_{\Lambda} Xe$,    .05\\
${\bf Cs}$&$^{116} Cs$,    .34&$^{187} Cs$,    .04&$^{117}_{\Lambda} Cs$,    .37&$^{188}_{\Lambda} Cs$,    .10\\
${\bf Ba}$&$^{113} Ba$,    .11&$^{190} Ba$,    .08&$^{114}_{\Lambda} Ba$,    .16&$^{191}_{\Lambda} Ba$,    .14\\
${\bf La}$&$^{121} La$,    .35&$^{193} La$,    .13&$^{122}_{\Lambda} La$,    .38&$^{196}_{\Lambda} La$,    .02\\
${\bf Ce}$&$^{118} Ce$,    .17&$^{198} Ce$,    .01&$^{119}_{\Lambda} Ce$,    .21&$^{199}_{\Lambda} Ce$,    .06\\
${\bf Pr}$&$^{126} Pr$,    .35&$^{201} Pr$,    .05&$^{127}_{\Lambda} Pr$,    .37&$^{202}_{\Lambda} Pr$,    .11\\
${\bf Nd}$&$^{123} Nd$,    .21&$^{204} Nd$,    .10&$^{124}_{\Lambda} Nd$,    .25&$^{205}_{\Lambda} Nd$,    .15\\
${\bf Pm}$&$^{131} Pm$,    .32&$^{207} Pm$,    .14&$^{132}_{\Lambda} Pm$,    .34&$^{210}_{\Lambda} Pm$,    .04\\
${\bf Sm}$&$^{128} Sm$,    .22&$^{212} Sm$,    .03&$^{129}_{\Lambda} Sm$,    .25&$^{213}_{\Lambda} Sm$,    .08\\
${\bf Eu}$&$^{136} Eu$,    .28&$^{215} Eu$,    .07&$^{137}_{\Lambda} Eu$,    .30&$^{216}_{\Lambda} Eu$,    .13\\
${\bf Gd}$&$^{133} Gd$,    .22&$^{218} Gd$,    .11&$^{134}_{\Lambda} Gd$,    .25&$^{221}_{\Lambda} Gd$,    .01\\
${\bf Tb}$&$^{141} Tb$,    .22&$^{223} Tb$,    .01&$^{142}_{\Lambda} Tb$,    .24&$^{224}_{\Lambda} Tb$,    .06\\
${\bf Dy}$&$^{138} Dy$,    .19&$^{226} Dy$,    .05&$^{139}_{\Lambda} Dy$,    .22&$^{227}_{\Lambda} Dy$,    .10\\
${\bf Ho}$&$^{146} Ho$,    .15&$^{229} Ho$,    .09&$^{147}_{\Lambda} Ho$,    .17&$^{230}_{\Lambda} Ho$,    .14\\
${\bf Er}$&$^{143} Er$,    .15&$^{232} Er$,    .13&$^{144}_{\Lambda} Er$,    .18&$^{235}_{\Lambda} Er$,    .04\\
${\bf Tm}$&$^{151} Tm$,    .07&$^{237} Tm$,    .03&$^{152}_{\Lambda} Tm$,    .08&$^{238}_{\Lambda} Tm$,    .08\\
${\bf Yb}$&$^{148} Yb$,    .10&$^{240} Yb$,    .07&$^{149}_{\Lambda} Yb$,    .12&$^{241}_{\Lambda} Yb$,    .12\\
${\bf Lu}$&$^{157} Lu$,    .31&$^{243} Lu$,    .11&$^{158}_{\Lambda} Lu$,    .32&$^{246}_{\Lambda} Lu$,    .02\\
${\bf Hf}$&$^{153} Hf$,    .03&$^{248} Hf$,    .02&$^{154}_{\Lambda} Hf$,    .05&$^{249}_{\Lambda} Hf$,    .06\\
${\bf Ta}$&$^{162} Ta$,    .20&$^{251} Ta$,    .06&$^{163}_{\Lambda} Ta$,    .21&$^{252}_{\Lambda} Ta$,    .10\\
${\bf W }$&$^{159} W $,    .27&$^{254} W $,    .10&$^{160}_{\Lambda} W $,    .29&$^{257}_{\Lambda} W $,    .01\\
${\bf Re}$&$^{167} Re$,    .08&$^{259} Re$,    .01&$^{168}_{\Lambda} Re$,    .08&$^{260}_{\Lambda} Re$,    .05\\
${\bf Os}$&$^{164} Os$,    .17&$^{262} Os$,    .04&$^{165}_{\Lambda} Os$,    .18&$^{263}_{\Lambda} Os$,    .09\\
${\bf Ir}$&$^{173} Ir$,    .25&$^{265} Ir$,    .08&$^{174}_{\Lambda} Ir$,    .25&$^{268}_{\Lambda} Ir$,    .00\\
${\bf Pt}$&$^{169} Pt$,    .07&$^{268} Pt$,    .12&$^{170}_{\Lambda} Pt$,    .08&$^{271}_{\Lambda} Pt$,    .04\\
${\bf Au}$&$^{178} Au$,    .11&$^{273} Au$,    .04&$^{179}_{\Lambda} Au$,    .12&$^{274}_{\Lambda} Au$,    .08\\
${\bf Hg}$&$^{175} Hg$,    .24&$^{276} Hg$,    .07&$^{176}_{\Lambda} Hg$,    .25&$^{277}_{\Lambda} Hg$,    .11\\
${\bf Tl}$&$^{184} Tl$,    .25&$^{279} Tl$,    .11&$^{185}_{\Lambda} Tl$,    .25&$^{282}_{\Lambda} Tl$,    .03\\
${\bf Pb}$&$^{180} Pb$,    .11&$^{284} Pb$,    .03&$^{181}_{\Lambda} Pb$,    .12&$^{285}_{\Lambda} Pb$,    .07\\
${\bf Bi}$&$^{189} Bi$,    .08&$^{287} Bi$,    .07&$^{190}_{\Lambda} Bi$,    .09&$^{288}_{\Lambda} Bi$,   .10\\  
\end{tabular}
\end{ruledtabular}
\end{table}

\begin{table}[htbp]
\caption{One-nucleon separation energies (in MeV) on driplines for each element with the lowest and highest number of bound neutrons in normal and $\Lambda \Lambda$-hypernuclei.}
\begin{ruledtabular}
\begin{tabular}{ccccc}
Symbol&Normal&Normal&Hyper&Hyper \\
&p-drip&n-drip&p-drip&n-drip \\ \hline
$Z$&$^AZ$, $S_p$&$^AZ$, $S_n$&$^AZ$, $S_p$&$^AZ$, $S_n$ \\ \hline
\hline
${\bf Li}$&$^{  5} Li$,   3.36&$^{ 11} Li$,    .58& - - , - - &$^{ 15}_{\Lambda\Lambda}Li$,    .60\\
${\bf Be}$&$^{  6} Be$,   1.11&$^{ 14} Be$,    .90&$^{  7}_{\Lambda\Lambda} Be$,   2.02&$^{ 18}       _{\Lambda\Lambda} Be$,    .59\\
${\bf B }$&$^{  8} B $,    .74&$^{ 17} B $,    .99&$^{  9}_{\Lambda\Lambda} B $,    .60&$^{ 21}       _{\Lambda\Lambda} B $,    .53\\
${\bf C }$&$^{  9} C $,    .17&$^{ 20} C $,   1.01&$^{ 10}_{\Lambda\Lambda} C $,    .28&$^{ 24}       _{\Lambda\Lambda} C $,    .49\\
${\bf N }$&$^{ 12} N $,   1.98&$^{ 23} N $,    .97&$^{ 13}_{\Lambda\Lambda} N $,   1.17&$^{ 27}       _{\Lambda\Lambda} N $,    .44\\
${\bf O }$&$^{ 13} O $,   1.98&$^{ 26} O $,    .94&$^{ 14}_{\Lambda\Lambda} O $,   1.40&$^{ 30}       _{\Lambda\Lambda} O $,    .42\\
${\bf F }$&$^{ 15} F $,    .09&$^{ 29} F $,    .89&$^{ 17}_{\Lambda\Lambda} F $,   1.43&$^{ 33}       _{\Lambda\Lambda} F $,    .40\\
${\bf Ne}$&$^{ 16} Ne$,    .64&$^{ 32} Ne$,    .87&$^{ 17}_{\Lambda\Lambda} Ne$,    .00&$^{ 36}       _{\Lambda\Lambda} Ne$,    .40\\
${\bf Na}$&$^{ 19} Na$,    .54&$^{ 35} Na$,    .84&$^{ 21}_{\Lambda\Lambda} Na$,   1.50&$^{ 39}       _{\Lambda\Lambda} Na$,    .41\\
${\bf Mg}$&$^{ 20} Mg$,   1.37&$^{ 38} Mg$,    .84&$^{ 21}_{\Lambda\Lambda} Mg$,    .63&$^{ 42}       _{\Lambda\Lambda} Mg$,    .44\\
${\bf Al}$&$^{ 23} Al$,    .71&$^{ 41} Al$,    .84&$^{ 25}_{\Lambda\Lambda} Al$,   1.46&$^{ 45}       _{\Lambda\Lambda} Al$,    .47\\
${\bf Si}$&$^{ 23} Si$,    .07&$^{ 44} Si$,    .86&$^{ 25}_{\Lambda\Lambda} Si$,   1.00&$^{ 48}       _{\Lambda\Lambda} Si$,    .51\\
${\bf P }$&$^{ 27} P $,    .74&$^{ 49} P $,    .04&$^{ 29}_{\Lambda\Lambda} P $,   1.35&$^{ 51}       _{\Lambda\Lambda} P $,    .56\\
${\bf S }$&$^{ 27} S $,    .46&$^{ 52} S $,    .12&$^{ 29}_{\Lambda\Lambda} S $,   1.22&$^{ 54}       _{\Lambda\Lambda} S $,    .61\\
${\bf Cl}$&$^{ 31} Cl$,    .68&$^{ 55} Cl$,    .21&$^{ 33}_{\Lambda\Lambda} Cl$,   1.20&$^{ 59}       _{\Lambda\Lambda} Cl$,    .00\\
${\bf Ar}$&$^{ 31} Ar$,    .69&$^{ 58} Ar$,    .29&$^{ 32}_{\Lambda\Lambda} Ar$,    .00&$^{ 62}       _{\Lambda\Lambda} Ar$,    .09\\
${\bf K }$&$^{ 35} K $,    .57&$^{ 61} K $,    .37&$^{ 37}_{\Lambda\Lambda} K $,   1.01&$^{ 65}       _{\Lambda\Lambda} K $,    .18\\
${\bf Ca}$&$^{ 35} Ca$,    .81&$^{ 64} Ca$,    .45&$^{ 36}_{\Lambda\Lambda} Ca$,    .13&$^{ 68}       _{\Lambda\Lambda} Ca$,    .26\\
${\bf Sc}$&$^{ 39} Sc$,    .42&$^{ 67} Sc$,    .53&$^{ 41}_{\Lambda\Lambda} Sc$,    .81&$^{ 71}       _{\Lambda\Lambda} Sc$,    .35\\
${\bf Ti}$&$^{ 39} Ti$,    .84&$^{ 72} Ti$,    .07&$^{ 40}_{\Lambda\Lambda} Ti$,    .19&$^{ 74}       _{\Lambda\Lambda} Ti$,    .43\\
${\bf V }$&$^{ 43} V $,    .24&$^{ 75} V $,    .17&$^{ 45}_{\Lambda\Lambda} V $,    .59&$^{ 79}       _{\Lambda\Lambda} V $,    .03\\
${\bf Cr}$&$^{ 43} Cr$,    .81&$^{ 78} Cr$,    .25&$^{ 44}_{\Lambda\Lambda} Cr$,    .18&$^{ 82}       _{\Lambda\Lambda} Cr$,    .12\\
${\bf Mn}$&$^{ 47} Mn$,    .04&$^{ 81} Mn$,    .34&$^{ 49}_{\Lambda\Lambda} Mn$,    .36&$^{ 85}       _{\Lambda\Lambda} Mn$,    .21\\
${\bf Fe}$&$^{ 47} Fe$,    .73&$^{ 84} Fe$,    .41&$^{ 48}_{\Lambda\Lambda} Fe$,    .13&$^{ 88}       _{\Lambda\Lambda} Fe$,    .28\\
${\bf Co}$&$^{ 52} Co$,    .76&$^{ 89} Co$,    .08&$^{ 53}_{\Lambda\Lambda} Co$,    .12&$^{ 91}       _{\Lambda\Lambda} Co$,    .36\\
${\bf Ni}$&$^{ 51} Ni$,    .61&$^{ 92} Ni$,    .16&$^{ 52}_{\Lambda\Lambda} Ni$,    .04&$^{ 96}       _{\Lambda\Lambda} Ni$,    .05\\
${\bf Cu}$&$^{ 56} Cu$,    .48&$^{ 95} Cu$,    .24&$^{ 58}_{\Lambda\Lambda} Cu$,    .72&$^{ 99}       _{\Lambda\Lambda} Cu$,    .13\\
${\bf Zn}$&$^{ 55} Zn$,    .47&$^{ 98} Zn$,    .31&$^{ 57}_{\Lambda\Lambda} Zn$,    .77&$^{102}       _{\Lambda\Lambda} Zn$,    .20\\
${\bf Ga}$&$^{ 60} Ga$,    .19&$^{103} Ga$,    .04&$^{ 62}_{\Lambda\Lambda} Ga$,    .41&$^{105}       _{\Lambda\Lambda} Ga$,    .28\\
${\bf Ge}$&$^{ 59} Ge$,    .30&$^{106} Ge$,    .11&$^{ 61}_{\Lambda\Lambda} Ge$,    .57&$^{110}       _{\Lambda\Lambda} Ge$,    .02\\
${\bf As}$&$^{ 65} As$,    .66&$^{109} As$,    .18&$^{ 66}_{\Lambda\Lambda} As$,    .11&$^{113}       _{\Lambda\Lambda} As$,    .09\\
${\bf Se}$&$^{ 63} Se$,    .11&$^{112} Se$,    .25&$^{ 65}_{\Lambda\Lambda} Se$,    .36&$^{116}       _{\Lambda\Lambda} Se$,    .16\\
${\bf Br}$&$^{ 69} Br$,    .33&$^{117} Br$,    .02&$^{ 71}_{\Lambda\Lambda} Br$,    .51&$^{119}       _{\Lambda\Lambda} Br$,    .23\\
${\bf Kr}$&$^{ 68} Kr$,    .62&$^{120} Kr$,    .08&$^{ 69}_{\Lambda\Lambda} Kr$,    .13&$^{124}       _{\Lambda\Lambda} Kr$,    .00\\
${\bf Rb}$&$^{ 73} Rb$,    .00&$^{123} Rb$,    .15&$^{ 75}_{\Lambda\Lambda} Rb$,    .17&$^{127}       _{\Lambda\Lambda} Rb$,    .07\\
${\bf Sr}$&$^{ 72} Sr$,    .36&$^{126} Sr$,    .21&$^{ 74}_{\Lambda\Lambda} Sr$,    .56&$^{130}       _{\Lambda\Lambda} Sr$,    .13\\
${\bf Y }$&$^{ 78} Y $,    .33&$^{131} Y $,    .01&$^{ 80}_{\Lambda\Lambda} Y $,    .47&$^{133}       _{\Lambda\Lambda} Y $,    .19\\
${\bf Zr}$&$^{ 76} Zr$,    .10&$^{134} Zr$,    .07&$^{ 78}_{\Lambda\Lambda} Zr$,    .28&$^{136}       _{\Lambda\Lambda} Zr$,    .25\\
${\bf Nb}$&$^{ 83} Nb$,    .58&$^{137} Nb$,    .13&$^{ 84}_{\Lambda\Lambda} Nb$,    .11&$^{141}       _{\Lambda\Lambda} Nb$,    .06\\
${\bf Mo}$&$^{ 81} Mo$,    .45&$^{140} Mo$,    .19&$^{ 82}_{\Lambda\Lambda} Mo$,    .00&$^{144}       _{\Lambda\Lambda} Mo$,    .12\\
${\bf Tc}$&$^{ 87} Tc$,    .20&$^{145} Tc$,    .01&$^{ 89}_{\Lambda\Lambda} Tc$,    .32&$^{147}       _{\Lambda\Lambda} Tc$,    .17\\ 
\end{tabular}
\end{ruledtabular}
\end{table}

\begin{table}[htbp]
\caption{One-nucleon separation energies (in MeV) on driplines for each element with the lowest and highest number of bound neutrons in normal and $\Lambda \Lambda$-hypernuclei.}
\begin{ruledtabular}
\begin{tabular}{ccccc}
Symbol&Normal&Normal&Hyper&Hyper \\
&p-drip&n-drip&p-drip&n-drip \\ \hline
$Z$&$^AZ$, $S_p$&$^AZ$, $S_n$&$^AZ$, $S_p$&$^AZ$, $S_n$ \\ \hline
\hline
${\bf Ru}$&$^{ 85} Ru$,    .14&$^{148} Ru$,    .07&$^{ 87}_{\Lambda\Lambda} Ru$,    .29&$^{150}       _{\Lambda\Lambda} Ru$,    .22\\
${\bf Rh}$&$^{ 92} Rh$,    .40&$^{151} Rh$,    .13&$^{ 94}_{\Lambda\Lambda} Rh$,    .50&$^{155}       _{\Lambda\Lambda} Rh$,    .06\\
${\bf Pd}$&$^{ 90} Pd$,    .39&$^{154} Pd$,    .18&$^{ 92}_{\Lambda\Lambda} Pd$,    .52&$^{158}       _{\Lambda\Lambda} Pd$,    .11\\
${\bf Ag}$&$^{ 96} Ag$,    .01&$^{159} Ag$,    .02&$^{ 98}_{\Lambda\Lambda} Ag$,    .11&$^{161}       _{\Lambda\Lambda} Ag$,    .16\\
${\bf Cd}$&$^{ 94} Cd$,    .06&$^{162} Cd$,    .07&$^{ 96}_{\Lambda\Lambda} Cd$,    .19&$^{166}       _{\Lambda\Lambda} Cd$,    .01\\
${\bf In}$&$^{101} In$,    .14&$^{165} In$,    .12&$^{103}_{\Lambda\Lambda} In$,    .23&$^{169}       _{\Lambda\Lambda} In$,    .06\\
${\bf Sn}$&$^{ 99} Sn$,    .25&$^{168} Sn$,    .17&$^{101}_{\Lambda\Lambda} Sn$,    .35&$^{172}       _{\Lambda\Lambda} Sn$,    .11\\
${\bf Sb}$&$^{106} Sb$,    .24&$^{173} Sb$,    .03&$^{108}_{\Lambda\Lambda} Sb$,    .31&$^{175}       _{\Lambda\Lambda} Sb$,    .16\\
${\bf Te}$&$^{104} Te$,    .38&$^{176} Te$,    .07&$^{105}_{\Lambda\Lambda} Te$,    .00&$^{180}       _{\Lambda\Lambda} Te$,    .02\\
${\bf I }$&$^{111} I $,    .30&$^{179} I $,    .13&$^{113}_{\Lambda\Lambda} I $,    .37&$^{183}       _{\Lambda\Lambda} I $,    .07\\
${\bf Xe}$&$^{108} Xe$,    .02&$^{182} Xe$,    .17&$^{110}_{\Lambda\Lambda} Xe$,    .11&$^{186}       _{\Lambda\Lambda} Xe$,    .11\\
${\bf Cs}$&$^{116} Cs$,    .34&$^{187} Cs$,    .04&$^{118}_{\Lambda\Lambda} Cs$,    .40&$^{189}       _{\Lambda\Lambda} Cs$,    .16\\
${\bf Ba}$&$^{113} Ba$,    .11&$^{190} Ba$,    .08&$^{115}_{\Lambda\Lambda} Ba$,    .20&$^{194}       _{\Lambda\Lambda} Ba$,    .03\\
${\bf La}$&$^{121} La$,    .35&$^{193} La$,    .13&$^{123}_{\Lambda\Lambda} La$,    .40&$^{197}       _{\Lambda\Lambda} La$,    .08\\
${\bf Ce}$&$^{118} Ce$,    .17&$^{198} Ce$,    .01&$^{120}_{\Lambda\Lambda} Ce$,    .25&$^{200}       _{\Lambda\Lambda} Ce$,    .12\\
${\bf Pr}$&$^{126} Pr$,    .35&$^{201} Pr$,    .05&$^{128}_{\Lambda\Lambda} Pr$,    .40&$^{205}       _{\Lambda\Lambda} Pr$,    .00\\
${\bf Nd}$&$^{123} Nd$,    .21&$^{204} Nd$,    .10&$^{125}_{\Lambda\Lambda} Nd$,    .28&$^{208}       _{\Lambda\Lambda} Nd$,    .05\\
${\bf Pm}$&$^{131} Pm$,    .32&$^{207} Pm$,    .14&$^{133}_{\Lambda\Lambda} Pm$,    .36&$^{211}       _{\Lambda\Lambda} Pm$,    .09\\
${\bf Sm}$&$^{128} Sm$,    .22&$^{212} Sm$,    .03&$^{130}_{\Lambda\Lambda} Sm$,    .28&$^{214}       _{\Lambda\Lambda} Sm$,    .14\\
${\bf Eu}$&$^{136} Eu$,    .28&$^{215} Eu$,    .07&$^{138}_{\Lambda\Lambda} Eu$,    .32&$^{219}       _{\Lambda\Lambda} Eu$,    .02\\
${\bf Gd}$&$^{133} Gd$,    .22&$^{218} Gd$,    .11&$^{135}_{\Lambda\Lambda} Gd$,    .27&$^{222}       _{\Lambda\Lambda} Gd$,    .07\\
${\bf Tb}$&$^{141} Tb$,    .22&$^{223} Tb$,    .01&$^{143}_{\Lambda\Lambda} Tb$,    .25&$^{225}       _{\Lambda\Lambda} Tb$,    .11\\
${\bf Dy}$&$^{138} Dy$,    .19&$^{226} Dy$,    .05&$^{140}_{\Lambda\Lambda} Dy$,    .24&$^{230}       _{\Lambda\Lambda} Dy$,    .00\\
${\bf Ho}$&$^{146} Ho$,    .15&$^{229} Ho$,    .09&$^{148}_{\Lambda\Lambda} Ho$,    .18&$^{233}       _{\Lambda\Lambda} Ho$,    .05\\
${\bf Er}$&$^{143} Er$,    .15&$^{232} Er$,    .13&$^{145}_{\Lambda\Lambda} Er$,    .20&$^{236}       _{\Lambda\Lambda} Er$,    .09\\
${\bf Tm}$&$^{151} Tm$,    .07&$^{237} Tm$,    .03&$^{153}_{\Lambda\Lambda} Tm$,    .10&$^{239}       _{\Lambda\Lambda} Tm$,    .13\\
${\bf Yb}$&$^{148} Yb$,    .10&$^{240} Yb$,    .07&$^{150}_{\Lambda\Lambda} Yb$,    .14&$^{244}       _{\Lambda\Lambda} Yb$,    .03\\
${\bf Lu}$&$^{157} Lu$,    .31&$^{243} Lu$,    .11&$^{158}_{\Lambda\Lambda} Lu$,    .00&$^{247}       _{\Lambda\Lambda} Lu$,    .07\\
${\bf Hf}$&$^{153} Hf$,    .03&$^{248} Hf$,    .02&$^{155}_{\Lambda\Lambda} Hf$,    .07&$^{250}       _{\Lambda\Lambda} Hf$,    .11\\
${\bf Ta}$&$^{162} Ta$,    .20&$^{251} Ta$,    .06&$^{164}_{\Lambda\Lambda} Ta$,    .22&$^{255}       _{\Lambda\Lambda} Ta$,    .02\\
${\bf W }$&$^{159} W $,    .27&$^{254} W $,    .10&$^{161}_{\Lambda\Lambda} W $,    .30&$^{258}       _{\Lambda\Lambda} W $,    .05\\
${\bf Re}$&$^{167} Re$,    .08&$^{259} Re$,    .01&$^{169}_{\Lambda\Lambda} Re$,    .09&$^{261}       _{\Lambda\Lambda} Re$,    .09\\
${\bf Os}$&$^{164} Os$,    .17&$^{262} Os$,    .04&$^{166}_{\Lambda\Lambda} Os$,    .20&$^{266}       _{\Lambda\Lambda} Os$,    .01\\
${\bf Ir}$&$^{173} Ir$,    .25&$^{265} Ir$,    .08&$^{175}_{\Lambda\Lambda} Ir$,    .26&$^{269}       _{\Lambda\Lambda} Ir$,    .05\\
${\bf Pt}$&$^{169} Pt$,    .07&$^{268} Pt$,    .12&$^{171}_{\Lambda\Lambda} Pt$,    .09&$^{272}       _{\Lambda\Lambda} Pt$,    .08\\
${\bf Au}$&$^{178} Au$,    .11&$^{273} Au$,    .04&$^{180}_{\Lambda\Lambda} Au$,    .12&$^{275}       _{\Lambda\Lambda} Au$,    .12\\
${\bf Hg}$&$^{175} Hg$,    .24&$^{276} Hg$,    .07&$^{177}_{\Lambda\Lambda} Hg$,    .26&$^{280}       _{\Lambda\Lambda} Hg$,    .04\\
${\bf Tl}$&$^{184} Tl$,    .25&$^{279} Tl$,    .11&$^{186}_{\Lambda\Lambda} Tl$,    .25&$^{283}       _{\Lambda\Lambda} Tl$,    .07\\
${\bf Pb}$&$^{180} Pb$,    .11&$^{284} Pb$,    .03&$^{182}_{\Lambda\Lambda} Pb$,    .12&$^{286}       _{\Lambda\Lambda} Pb$,    .11\\
${\bf Bi}$&$^{189} Bi$,    .08&$^{287} Bi$,    .07&$^{191}_{\Lambda\Lambda} Bi$,    .09&$^{291}       _{\Lambda\Lambda} Bi$,    .03\\ 
\end{tabular}
\end{ruledtabular}
\end{table}

\noindent
\section{Lambda hyperonic effect on the driplines}

The effect of addition of a single $\Lambda$ and $\Lambda\Lambda$ in a non-strange normal nucleus can be seen through the one-neutron and one-proton separation energies tabulated in Tables 1,2 and Tables 3,4 respectively.

Tables 1,2, containing the dripline nuclei of $\Lambda$-hypernuclei, delineate that the $\Lambda$-hyperonic effects are more prominent on the neutron dripline than the proton dripline. The proton dripline of single $\Lambda$-hypernuclei are shifted only for few light nuclei (e.g.  N, O, Mg etc.) and as the number of nucleon increases the dripline merges to normal proton dripline possibly due to the weak $\Lambda$-nucleon \cite{Taka02} coupling effect on static mean field potential. Some normal neutron drip nuclei like $^{44}Si$ (N=30) and $^{193}La$ (N=136) change to  $^{47}_\Lambda Si$ (N=32) and $^{196}_\Lambda La$ (N=138) respectively due to the addition of single $\Lambda$ to the core so that the neutron dripline of $\Lambda$-hypernuclei shifts away from the stability line.

Similar picture emerges from Table 3 and 4 for the proton and neutron dripline of $\Lambda \Lambda$-hypernuclei. In this case relatively larger effect on proton and neutron dripline nuclei occurs due to the enhanced hyperon-nucleon coupling effect. The proton drip line for $\Lambda \Lambda$-hypernuclei moves away from the stability line for the light and medium heavy nuclei (e.g. $Be$ to $O, Ne, Mg, Ar, Ca, Ti, Cr, Fe, Co, Ni, As,$ $ Kr, Nb, Mo, Te, Lu$ etc.). The neutron dripline of $\Lambda\Lambda$-hypernuclei is shifted more prominently for almost all nuclei  from light to heavy in comparison to the normal neutron dripline. For example, addition of a $\Lambda\Lambda$ in normal neutron drip nuclei $^{151}$Rh, $^{154}$Pd, $^{165}$In and $^{168}$Sn leads to more neutron-rich drip line hypernuclei $_{\Lambda\Lambda}^{155}Rh$ (instead of $_{\Lambda\Lambda}^{153}Rh$), $_{\Lambda\Lambda}^{158}Pd$ (instead of $_{\Lambda\Lambda}^{156}Pd$), $_{\Lambda\Lambda}^{169}In$ (instead of $_{\Lambda\Lambda}^{167}In$) and $_{\Lambda\Lambda}^{172}Sn$ (instead of $_{\Lambda\Lambda}^{170}Sn$) respectively. \\

It is interesting to note that although for some cases the neutron number of some neutron dripline $\Lambda$ hypernuclei and $\Lambda\Lambda$ hypernuclei are the same, their neutron separation energies are different. For example,  $_{\Lambda}^{47}Si$ (Table 1) and $_{\Lambda\Lambda}^{48}Si$ (Table 3) hypernuclei have the same neutron number (N=32), but the values of $S_n$ are 0.242 MeV and 0.513 MeV repectively. It  indicates that addition of single and double lambda hyperons to normal nuclei modify the mean field differently.

Drip lines for single $\Lambda$ and  $\Lambda\Lambda$ hypernuclei delineate interesting changes on both sides indicating effect of change of  nuclear potential depth in hypernuclei compared to the normal nuclei. This observation may have important consequences in case of the astrophysical objects where the dense strangeness-rich matter may exist in the core of compact stars or for the evolution of the strange stars.

\noindent
\section{Summary}

The mass formula BWMH, applicable to both non-strange normal nuclei and strange hypernuclei, has been employed to evaluate the neutron and proton separation energies from  non-strange normal nuclei as well as those hypernuclei which contain either one  $\Lambda$ hyperon or,  two  $\Lambda$  hyperons. 
It is found that the neutron and proton drip lines for non-strange normal nuclei and hypernuclei are not the same. Addition of a $\Lambda$ (or, $\Lambda\Lambda$) does not merely act as an addition of a neutral particle like neutron, instead it affects different nuclei differently and many exotic nuclei beyond the normal drip lines are possible.  Tables 1-4 illustrate distinct proof of this interesting new finding. 

Normal neutron drip line nuclei can not hold any further neutron(s) added to them. But, they can hold additional hyperons which is possible due to strong hyperon-nucleon interaction. Interestingly, addition of hyperon can sometimes help the nuclei to hold more neutrons than possible in normal nuclei. For example, Table 1 shows that addition of $\Lambda$ in light neutron-drip line nuclei (Li, Be, B etc.) has little effect on the total neutron number content, where as some heavier hypernuclei (Sc, Cu, Zn etc.) can retain more neutrons than their corresponding normal drip line nuclei. Again, on the proton drip side, a few lambda hypernuclei are formed more neutron deficient than the normal ones. Thus both more neutron-rich or, more neutron-deficient nuclei beyond the normal drip lines are possible for hypernuclei with a single $\Lambda$.

Similar phenomenon is found for the $\Lambda\Lambda$ hypernuclei. Table 3-4 show that on the proton drip side, $^{39}$Ti (N=17) is a normal proton drip nucleus, but after adding  $\Lambda\Lambda$, the proton drip hypernucleus is $_{\Lambda\Lambda}^{40}Ti$ (N=16), not $_{\Lambda\Lambda}^{41}Ti$ (N=17). On the other hand, for normal proton drip nucleus $^{85}$Ru, addition of $\Lambda\Lambda$ makes the proton drip hypernucleus $_{\Lambda\Lambda}^{87}Ru$, retaining the same neutron number (N=41). On the neutron drip side we see that addition of  $\Lambda\Lambda$ to the core of normal dripline nuclei sometimes increases their capacity to bind more neutrons. 
For example, the normal neutron drip nuclei $^{23}N$ (N=16) and $^{232}Er$ (N=164) get shifted to more neutron rich nuclei $_{\Lambda\Lambda}^{27}N$ (N=18) and $_{\Lambda\Lambda}^{236}Er$ (N=166) respectively.
On the other hand, addition of double lambda to normal neutron drip nuclei $^{49}P$, $^{52}S$, $^{89}Co$ makes neutron drip hypernuclei $_{\Lambda\Lambda}^{51}P$, $_{\Lambda\Lambda}^{54}S$, $_{\Lambda\Lambda}^{91}Co$ with the same neutron numbers. \\

 In summary, addition of $\Lambda$ or, $\Lambda\Lambda$ hyperon(s) to a normal nucleus is found to cause shifts of the neutron and proton driplines from their normal limits. But, this shift is not uniform along the entire nuclear chart. The non-uniform shift of hypernuclear driplines from the normal driplines is possibly a manifestation of intricate changes in mean-field as well as nucleon-hyperon or, hyperon-hyperon interactions in nuclear matter of different neutron-proton asymmetry. These aspects as well as any structure effect need to be investigated in a microscopic framework which is beyond the scope of this work. Possible existence of several bound exotic nuclei, similar to $_\Lambda^{10}Li$  for which the normal nuclei with same mass number are unbound, are suggested near the drip lines. In this context, experimental  search for the bound exotic $_\Lambda^{13}O$ ($S_\Lambda$ = 14.1 MeV) and $_{\Lambda\Lambda}^{15}Li$  ($S_{\Lambda\Lambda}$ = 32.9 MeV) hypernuclei would be interesting as their cores $^{12}O$ and $^{13}Li$ are beyond the normal drip lines.

\end{document}